\documentclass[12pt]{article}
\usepackage{graphics}
\usepackage{amsmath}
\usepackage{amssymb}
\usepackage{graphicx}
\usepackage{makeidx}
\usepackage{mathrsfs}
\usepackage{amsfonts}
\usepackage[mathscr]{eucal}
\usepackage{amsmath}

\usepackage{graphics}

\DeclareGraphicsExtensions{.eps,.pdf,.jpg,.png}

\def\e{{\rm e}}
\def\e{\hbox{e}}

\def\ds{\displaystyle}
\def\RR{\vbox {\hbox to 8.9pt {I\hskip-2.1pt R\hfil}}}

\def\q{\quad}  
\def\cen{\centerline}
\def \rec#1{{\frac{1}{#1}}}
\def\pni{\par\noindent}
\def\vsh{\smallskip}

\def\vsp{\vsh\pni} 

\DeclareGraphicsExtensions{.eps,.pdf,.jpg,.png}

\begin{document}

\cen{{\bf FRACALMO PRE-PRINT: \ http://www.fracalmo.org}}
%
\vsh
\hrule
\vskip 0.50truecm
\font\title=cmbx12 scaled\magstep2
\font\bfs=cmbx12 scaled\magstep1
\font\little=cmr10
\begin{center}
{\title On the viscoelastic characterization of }
\\[0.25truecm]
{\title the Jeffreys--Lomnitz law of creep}
 \\  [0.25truecm]
 Francesco MAINARDI$^{(1)}$ and
Giorgio SPADA$^{(2)}$
\\  [0.25truecm]
$\null^{(1)}$ {\little Department of Physics, University of Bologna, and INFN}
\\
{\little Via Irnerio 46, I-40126 Bologna, Italy}
\\{\little Corresponding Author.   E-mail: francesco.mainardi@bo.infn.it}
\\[0.25truecm]
$\null^{(2)}$ {\little Dipartimento di Scienze di Base e Fondamenti,
 University of Urbino,}
 \\ {\little Via Santa Chiara 27, I-61029 Urbino, Italy}
 \\{\little E-mail: giorgio.spada@gmail.com}
\\[0.25truecm]
{\bf Paper published in Rheologica Acta, Vol. 51 (2012), pp. 783-791}
\end{center}
\begin{abstract}
\noindent
In 1958 Jeffreys proposed a  power law of creep, generalizing the logarithmic law
 earlier introduced by  Lomnitz, to broaden  the geophysical applications
to  fluid-like materials including igneous rocks.
This generalized law, however, can be applied also to solid-like  viscoelastic materials.
We revisit the Jeffreys-Lomnitz law of creep 
by  allowing its power law exponent  $\alpha$, usually limited to the range $0\le \alpha\le 1$ 
 to all negative values.
This  is consistent with the linear theory of viscoelasticity  because
  the creep function still remains a Bernstein function, that is positive with a completely monotone
  derivative, with a related spectrum of retardation times. 
  The complete range  $\alpha \le 1$ yields a continuous transition 
  from a  Hooke elastic solid with no creep $\left(\alpha \!\to\! -\infty\right)$ to
   a Maxwell fluid with linear creep 
  $\left(\alpha \!=\!1\right)$ passing through   the  Lomnitz viscoelastic body 
  with logarithmic creep $\left(\alpha\! =0\right)$, which
  separates  solid-like from fluid-like behaviors. 
  Furthermore, we  numerically compute the relaxation modulus
  and provide the analytical expression of the  spectrum of retardation times
  corresponding to the Jeffreys-Lomnitz creep law extended to all $\alpha \le 1$.
   \end{abstract}
%

\vsp
\newpage
\section{Introduction}
 According to the linear uni-axial theory of viscoelasticity, 
 a material body can be viewed as a linear system 
 where either  stress $\sigma(t)$ or strain $\epsilon(t)$ have the role of excitation 
 function (input) and response function (output).
 From  both the experimental and theoretical viewpoints,  
 a central role is played by the {\it creep} and {\it relaxation tests}, 
 where the inputs are a step-wise  stress and strain, respectively.  
Introducing the Heaviside step function $\Theta(t)$,
in a {creep test}   $\sigma(t)=\sigma_{0}\Theta(t)$, 
conversely in a {relaxation test}  $\epsilon(t)=\epsilon_{0}\Theta(t)$, 
with constants $\sigma_0$ and $\epsilon_{0}$.
The corresponding outputs are characterized by the so-called {\it time-dependent material functions},
the  {\it creep compliance} $J(t)=\epsilon(t)/\sigma_{0}$ and 
 the {\it relaxation modulus}  $G(t)=\sigma(t)/\epsilon_{0}$, respectively. 
 From experimental evidence, $J(t)$ is non-decreasing and non-negative,
while $G(t)$ is non-increasing and non-negative.

In Earth rheology, transient creep is often described by 
 the Jeffreys-Lomnitz law. 
This law of creep is defined in terms of  four parameters as follows
\begin{equation}
\label{eq:(M1)} 
 J(t)= J_0 [1 + q \psi(t)] \,, \quad  t\ge 0\,,
 \end{equation}
 where $J_0$  is the unrelaxed compliance, $q$ is a  positive dimensionless material constant, and
  \begin{equation}
\label{eq:(M4)} 
\psi(t)= \frac{\left(1 + \displaystyle \frac{t}{\tau_0} \right)^\alpha-1}{\alpha}\,, \quad
0\le \alpha \le 1\,, \quad  t\ge 0\,.
\end{equation}
Here, $\tau_0$ is a characteristic time and  the exponent $\alpha$  essentially determines 
the rheological behaviour of the creep law. 
 By its own definition, 
 $\psi(t)$ is a positive increasing function of time
  that we refer to as the dimensionless  {\it Jeffreys-Lomnitz creep law}.
\vsp
Creep law (\ref{eq:(M1)}) was proposed by Sir Harold Jeffreys in 1958 \cite{Jeffreys_GJRAS58}
(see also \cite{Jeffreys_EARTH76} and \cite{Jeffreys-Crampin_MNRAS60,Jeffreys-Crampin_MNRAS70}),
 to generalize, via the parameter $\alpha$, the logarithmic law 
  \begin{equation}
\label{eq:(M3)} 
\psi(t) = \log \left(1 + \frac{t}{\tau_0}\right) \,, \quad t\ge 0\,.
\end{equation} introduced by Lomnitz
 in 1956 \cite {Lomnitz_JG56}  
 to describe flow in igneous rocks.
This law was also employed by Lomnitz to account for the damping of the 
Earth's free nutation (Chandler wobble) and of seismic S-waves, 
see Lomnitz (1957, 1962) \cite{Lomnitz_JAP57,Lomnitz_JGR62}.
  \vsp
As a matter of fact, for $\alpha=0$, the Jeffreys law
reduces  (in the limit) to the Lomnitz logarithmic law; while for $\alpha=1$,
it reduces to  the simple linear law of the Maxwell viscoelastic body.
According to Jeffreys,  Eq. (\ref{eq:(M4)}) 
 better fits the data of creep and dissipation in the Earth
for seismological purposes. The Jeffreys-Lomnitz law  is indeed well recognized in rheology of
rocks, in view of its property of interpolating creep data between a logarithmic and a linear
law, see \textit{e.g.} Jeffreys \cite{Jeffreys_EARTH76} and Ranalli \cite{Ranalli_BOOK87}.
 \vsp
 Although in the original papers the exponent $\alpha$ was not explicitly limited to positive values,
 the applications with $\alpha<0 $ were only consequently introduced in Earth rheology.   
  We note that the late Professor Ellis Strick in his 1984 paper \cite{Strick_JGR84}
had already extended the Jeffreys-Lomnitz law of creep allowing the parameter $\alpha$
in the range $-1 \le \alpha\le +1$, by introducing  parameter $s:= 1-\alpha$ 
($0\le s \le 2$). Furthermore, he was also interested in the representation of the
extended Jeffreys-Lomnitz law of creep in terms of a suitable ladder network of
springs and dashpots. In his work, Strick was motivated by some experimental observations
suggesting negative values of the exponent $\alpha$. At the time we started  editing  the present
paper, we were not aware of any reaction of the geophysical community to his
results. However, later we noted some published works, see \textit{e.g.}
 Crough \& Burford (1977) \cite{Crough-Burford_T77},
Spencer (1981) \cite{Spencer_JGR81},
Wesson (1988)  \cite{Wesson_JGR88},  Darby \& Smith (1990) \cite{Darby-Smith_GJI90}, where the
Jeffreys-Lomnitz law of creep was applied with $\alpha<0$ to fit experimental
geophysical data.  
In these papers, however, we did not find any attempt to characterize  the viscoelastic properties   
of the  Jeffreys-Lomnitz law in the whole range
$-\infty<\alpha \le 1$. In particular, the  relaxation modulus and the  retardation spectrum 
associated to this creep law have never been investigated up to now.
 Of course, in the trivial case  $\alpha=1$, we recover the Maxwell body  for which
the retardation spectrum  vanishes
and the relaxation modulus  is exponentially decreasing.
 \vsp
The purpose of this paper is to  characterize the  Jeffreys-Lomnitz law of creep
 consistently with the linear theory of viscoelasticity. 
In order to meet this goal, we find it convenient for the readers to recall in Section 2 the essential
mathematical notions of linear viscoelasticity necessary to understand our analysis. 
Then, in Section 3,   we consider  this law of creep allowing the parameter $\alpha$
in the whole range $-\infty < \alpha \le 1$.  
 Here, for some selected values of $\alpha$ in the  range $-2 \le \alpha \le 1$, we show plots 
of the analytically  determined creep function  and of the corresponding relaxation function,
numerically computed   by solving an integral Volterra equation of the second kind.
In Section 4, we  analytically derive the  spectrum of retardation times corresponding
to the extended creep law by using the tools of Laplace transforms, 
and we  exhibit plots for the above selected
values of $\alpha$.
 In Section 5, we draw our  conclusions and final remarks.
We have assumed  that the reader 
is familiar 
with the main properties and tables of 
Laplace transforms so we do not recall them. However, we devote the Appendix 
 to the Post-Widder  formula of Laplace inversion on the real axis, which yields a
complementary approach to our  derivation of the retardation spectrum.

\section{Essentials of linear viscoelasticity}
The basic principles of linear viscoelasticity are illustrated in well known treatises
\cite{Gross_BOOK53, Bland_BOOK60, Ferry_BOOK80, Christensen_BOOK82, Pipkin_BOOK86, Tschoegl_BOOK89,
Lin_BOOK2011}.
Here we mostly follow the notation of Chapter 2 of the recent book of Mainardi~\cite{Mainardi_BOOK10}, 
where the approaches by Gross~\cite{Gross_BOOK53}, Pipkin~\cite{Pipkin_BOOK86}
and Tschoegl~\cite{Tschoegl_BOOK89}
have been mainly adopted and revisited. Without loss of generality, the one-dimensional  theory will be
employed throughout.
\vsp
Assuming causal and differentiable histories of stress and strain (i.e., $\sigma(t)$ and $\epsilon(t)$
vanish for all times $t < 0$ and are differentiable for $t>0$), the time-dependent material functions
formerly introduced, $J(t)$ and $G(t)$, can be used to 
express the general stress-strain relationship as
\begin{equation}
\label{eq:(2.5a)} 
\epsilon (t)=  
 \sigma (0^+)\, J(t) + \int_0^t  \!\! J(t-\tau)\,
 \dot\sigma (\tau ) \, d\tau 
= \sigma (0^+)\, J(t) + J(t)\ast \dot \sigma(t),
\end{equation}
or, equivalently
\begin{equation}
\label{eq:(2.5b)} 
\sigma  (t) =
\epsilon(0^+)\, G(t) + \int_0^t  \!\! G(t-\tau)\,
 \dot \epsilon  (\tau ) \, d\tau
 =  \epsilon(0^+)\, G(t) + G(t)\ast \dot \epsilon(t),
\end{equation}
where dots and $\ast$ denote time-differentiation and time-convolution, respectively. 
\vsp
The limiting values of the time-dependent material functions for $t \to 0^+$ and $t \to +\infty$ describe the
instantaneous  and equilibrium  behaviour of the viscoelastic
body, respectively. The instantaneous $J_{0}$ and the equilibrium compliance
$J_{\infty}$ are defined by $J_0 \equiv J(0^+)$ and $J_\infty \equiv J(+\infty)$,
 while the instantaneous and equilibrium  modulus are $G_0 \equiv G(0^+)$ and $ G_\infty \equiv G(+\infty)$,
respectively.
\vsp
Taking the Laplace transform of Eqs. (\ref{eq:(2.5a)}) and (\ref{eq:(2.5b)}),
the following {\it reciprocity relation} is obtained
  \begin{equation}
\label{eq:(2.8)} 
  s\, \widetilde J(s)  = \rec{ s\, \widetilde G(s)} \quad \textrm{or} \quad
    \widetilde J(s) \, \widetilde G(s) = \rec{s^2},
 \,\,
 \end{equation}
where $s$ is the Laplace complex variable and the tilde denotes transformed
quantities\footnote{For the most relevant properties of the Laplace transform, 
we refer, e.g,. to the Appendix A3 (Eq. 32) of the well-known treatise by Tschoegl, see \cite{Tschoegl_BOOK89}
pp. 560--570.
In the following we use the 
notation $f(t)\,\div\, \widetilde f(s)$
to denote  the juxtaposition of an original  function $f(t)$
with its Laplace transform $\widetilde f(s)$.}.   
 Applying the limiting theorems of the Laplace transform
in the first  of Eqs. (\ref{eq:(2.8)}) provides
\begin{equation}
\label{eq:(2.14)} 
J_0 = \rec {G_0} \quad  \textrm{and} \quad J_\infty = \rec{G_\infty} \,,
\end{equation}
which allow us for a classification of the viscoelastic bodies into four types, according to the values of moduli 
and compliances (Table~\ref{table}).
\vskip 0.25truecm
\begin{table}[h!]
\caption{Classification of the four types of viscoelastic bodies according to values of
$J_{0}, J_{\infty}, G_{0}$ and $G_{\infty}$ (conventionally, $1/0=\infty$).}
\begin{center}
\begin{tabular}{|c||c|c||c|c|}
\hline
Type & $J_0$ & $J_\infty$ & $G_0$ & $G_\infty$ \\
\hline
I   & $>0$ & $<\infty$ & $<\infty$ & $>0$ \\
II  & $>0$ & $=\infty$ & $<\infty$ & $=0$ \\
III & $=0$ & $<\infty$ & $=\infty$ & $>0$ \\
IV  & $=0$ & $=\infty$ & $=\infty$ & $=0$ \\
\hline
\end{tabular}  
\label{table}
\end{center}
\end{table}
\vsp
Type I bodies exhibit both instantaneous and equilibrium elasticity, so they behave similarly to an elastic (Hooke) body
for sufficiently short and long times. Bodies of type II and IV show a complete stress relaxation (at constant strain)
since $G_\infty =0$ and an infinite strain creep (at constant stress) since $J_\infty = \infty,$,
so they do not exhibit equilibrium elasticity.
Finally, bodies of type III and IV do not present instantaneous elasticity since $J_0 = 0\,$
($G_0 =\infty$), hence they will not be considered in this study.
By the Laplace inversion of the second of  Eqs. (\ref{eq:(2.8)}), we obtain
the interrelation between the material functions in the time domain
\begin{equation}
\label{eq:(2.9)} 
J(t)   \,*\, G(t) := \int_0^t \!\! J(t-t')\,G(t')\,dt'
          = t\,, \quad t\ge 0,
\end{equation}
which will be useful in the following to obtain, by numerical methods,
$G(t)$ from the knowledge of $J(t)$ in the case of the (generalized) Jeffreys-Lomnitz law
of creep. In fact, if $J_0 > 0$ (types I and II),  Eq. (\ref{eq:(2.9)}) can be
re-written as a Volterra integral equation of the second kind
\begin{equation}
\label{eq:(2.12a)} 
G(t) = \frac{1}{J_0} - \frac{1}{J_0}\, \int_0^t\, \dot J(t-t' )\,
G(t')\, d t'\,,
\end{equation}
with $G(t)$ unknown. Of course, a similar equation can be established for
$J(t)$, provided that $G(t)$ is known and $G_{0}< \infty$ (types I and II). 
\vsp
Another relevant consequence of Eq. (\ref{eq:(2.9)}) is
\begin{equation}
\label{eq:(JG)} 
J(t)\, G(t) \le 1\,, \quad t\ge 0\,,
\end{equation}
where the equality holds in the limiting cases $t \to 0^+$ and $t\to +\infty$.
The results in Eqs. (\ref{eq:(2.12a)}), (\ref{eq:(JG)}) are discussed in \cite{Pipkin_BOOK86}
and recently revisited in \cite{Mainardi_BOOK10}.
\vsp
We remark  the restrictive conditions related
to the monotonicity of the time-dependent material functions.
According to Gross \cite{Gross_BOOK53}, their general expressions consistent
with experimental observations are as follows:
\begin{equation}
\label{eq:(2.28)} 
\begin{cases}
   J(t) =  J_0 +  \displaystyle \int_{0}^{\infty}
  R_\epsilon (\tau)\, \left( 1-\e^{\ds-t/\tau}\right)\, d\tau
        + J_+\, t \,,   \\
   G(t)  =  G_\infty +  \displaystyle \int_{0}^{\infty}
  R_\sigma  (\tau)\, \e^{\ds-t/\tau}\, d\tau
        + G_-\, \delta (t)\,,
\end{cases}
\end{equation}
where  the non-negative functions $R_\epsilon(\tau)$  and $R_\sigma(\tau)$ 
denote the {\it retardation spectrum} and the {\it relaxation  spectrum}, respectively.
For the classical mechanical models,
obtained by series and parallel combinations of a finite number of springs and dashpots 
\cite{Mainardi_BOOK10}, the spectra are discrete.
 \vsp
It is convenient to consider separately, for $t\ge 0$, those terms of
(\ref{eq:(2.28)}) deriving from a continuous  spectrum
 \begin{equation}
\label{eq:(2.30)} 
\begin{cases}
 J_\tau(t) := {\ds
   \int_{0}^{\infty}
  R_\epsilon (\tau)\, \left( 1-\e^{\ds-t/\tau}\right)\, d\tau} \,,\\
  G_\tau(t) := {\ds \int_{0}^{\infty}
  R_\sigma (\tau)\, \e^{\ds-t/\tau}\, d\tau}   \,,
\end{cases}
\end{equation}
where $J_\tau(t)$  (the {\it creep function with spectrum}) is a non-decreasing, non-negative function  with limiting values
$J_\tau(0^+)=0 $ and  $J_\tau(+\infty)\le \infty$,
whereas $G_\tau(t)$ (the {\it relaxation function with spectrum})
is a non-increasing, non-negative function,  with  $G_\tau(0^+) \le \infty$ and  $G_\tau(+\infty)= 0$.
The  integrability  on $\mathbb{R}^{+}$ of $R_\epsilon (\tau)$ and $R_\sigma(\tau)$
is ensured if $J_\tau(+\infty)< \infty$. 
\vsp
We also note that, consistently with the spectral representations (\ref{eq:(2.30)}), we have
$(-1)^n \, J_{\tau}^{(n)}(t) \le 0$ and $(-1)^n G_{\tau}^{(n)}(t) \ge 0$,  where the superscript denotes the
$n$--th time derivative.
From a mathematical standpoint, these conditions are equivalent to require
that $J_\tau(t)$ and $G_\tau(t)$ are {\it Bernstein} and {\it complete monotone functions}, respectively.
For details,  we refer to specialized mathematical treatises, 
Berg and Forst \cite{Berg-Forst_BOOK75}, Gripenpeg et al. \cite{Gripenberg_BOOK90},
and Schilling et al. \cite{Schilling_BOOK10}. These properties 
have been investigated by several authors, including Molinari \cite{Molinari_75}, Del Piero
and Deseri \cite{DelPiero-Deseri_QAM95} and, more recently, by Hanyga  \cite{Hanyga_RHEOACTA05}.
\vsp
The determination of the {\it time-spectral functions}  from the knowledge of
the time-dependent material
 functions is a problem that can be formally solved through a method,
outlined by Gross \cite{Gross_BOOK53}, based  on {\it Laplace transform pairs}.
For this purpose we introduce
 the {\it frequency-spectral functions}  as
\begin{equation}
\label{eq:(2.32)} 
 S_\epsilon(\gamma) :=
             \frac{R_\epsilon (1/\gamma)}{\gamma^2}\,,\q
 S_\sigma(\gamma):=
            \frac{R_\sigma  (1/\gamma)}{\gamma^2}\,,
 \end{equation}
where $\gamma =1/\tau $  denotes a retardation or relaxation frequency. Noting that
$R_\epsilon (\tau)\,d\tau= S_\epsilon (\gamma)\,d\gamma$ and $R_\sigma  (\tau)\,d\tau=
S_\sigma(\gamma)\,d\gamma$,
time differentiation of (\ref{eq:(2.30)}) yields
 \begin{equation}
\label{eq:(2.34)} 
\begin{cases}
+~ \dot J_\tau(t) = {\ds  \int_0^\infty \!\!
   \frac{R_\epsilon (\tau )}{\tau}\, \e^{-\ds t/\tau }\,d\tau
      = \int_0^\infty \!\!  \gamma \, S_\epsilon (\gamma )\,
  \e^{-\ds t\gamma}\,d\gamma}
      \,, \\ \\
 -\dot G_\tau (t)  =
 { \ds \int_0^\infty \!\!
       \frac{R_\sigma (\tau)}{\tau}\, \e^{-\ds t/\tau }\,d\tau
  = \int_0^\infty \!\! \gamma \,  S_\sigma (\gamma )\,
  \e^{-\ds t\gamma}\,d\gamma}
 \,,
\end{cases}
\end{equation}
showing that $\gamma \,S_\epsilon (\gamma )$ and $\, \gamma \,S_\sigma (\gamma )$ can be viewed as
the inverse Laplace transforms of $\dot J_\tau(t)$ and $-\dot G_\tau(t)\,,$ 
respectively, where $t$ is now considered the Laplace transform variable instead of the usual $s$. 
Thus, adopting the connective symbol $\div$ for Laplace transform pairs where in the LHS we put
the original function and in the R.H.S. its Laplace transform, we have
\begin{equation}
\label{eq:(2.35)} 
\begin{cases}
 +  \gamma \,S_\epsilon (\gamma) =
 {\ds \frac{R_\epsilon (1/\gamma)}{\gamma}
   \,\div\, \dot J_\tau(t)}\,, \\ \\
 - \gamma \,S_\sigma (\gamma) =
 {\ds \frac{R_\sigma  (1/\gamma)}{\gamma}
 \,\div\,  \dot G_\tau(t)}\,.
\end{cases}
\end{equation}
Consequently, when $J_\tau(t)$ and $G_\tau(t)$ are known analytically, the
corresponding frequency-spectral functions can be derived by standard
methods for the inversion of Laplace transforms; then, from Eq.
(\ref{eq:(2.32)}), the time-spectral functions are easily obtained.
  \vsp


\section{The  extended Jeffreys-Lomnitz laws of creep and relaxation}
\begin{figure}[h!]
\begin{center}
\includegraphics[width=0.6\textwidth,angle=-90.]{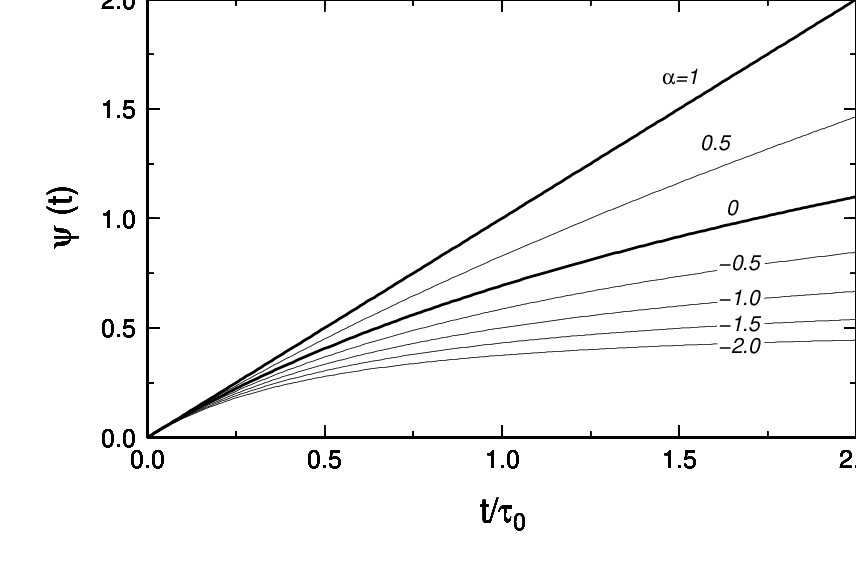}
\includegraphics[width=0.6\textwidth,angle=-90.]{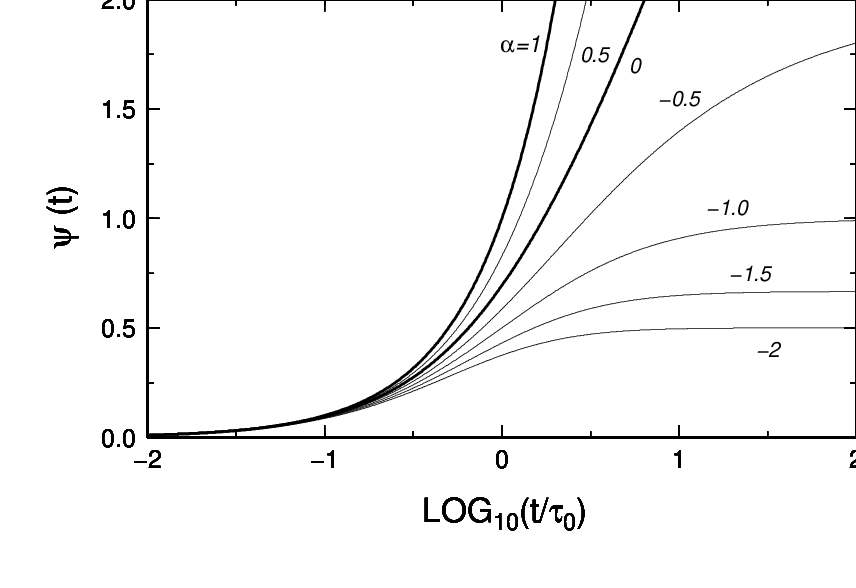}
\caption{
The  creep function $\psi(t)$
for $\alpha = 1, 0.5, 0, -0.5, -1, -1.5, -2$
versus dimensionless time $t/\tau_0$,
adopting  a linear scale (top) and a logarithmic scale (bottom).}
\label{fig1}
\end{center}
\end{figure}

\begin{figure}[h!]
\begin{center}
\includegraphics[width=0.6\textwidth,angle=-90.]{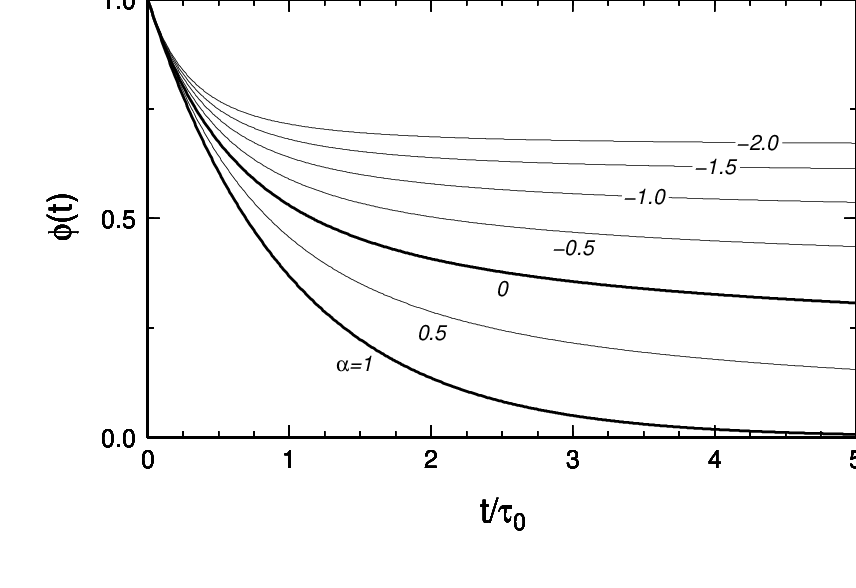}
\includegraphics[width=0.6\textwidth,angle=-90.]{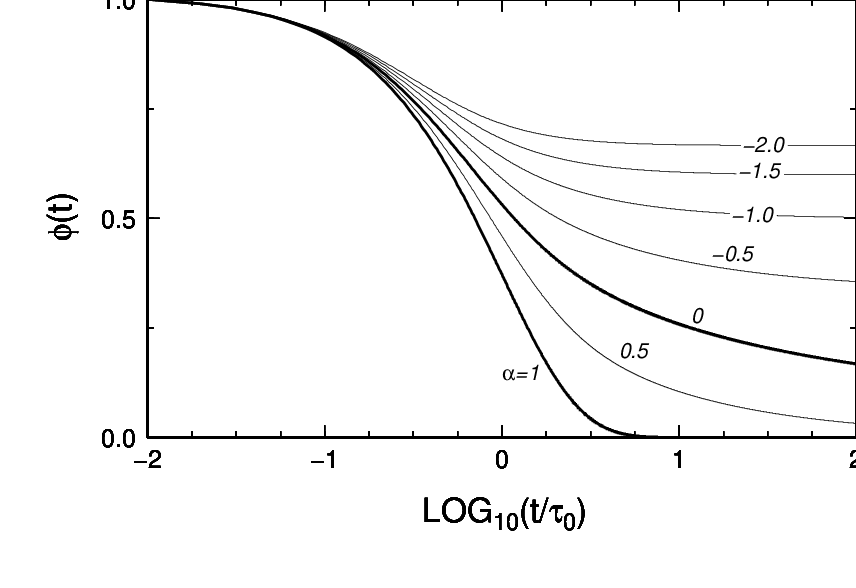}
\caption{
The   relaxation function $\phi(t)$
for $\alpha = 1, 0.5, 0, -0.5, -1, -1.5, -2$
versus dimensionless time $t/\tau_0$,
adopting  a linear scale (top) and a logarithmic scale (bottom).}
\label{fig2}
\end{center}
\end{figure}

In this Section we revisit the Jeffreys-Lomnitz law of creep by extending the range
of variability of the exponent $\alpha$ to any negative value  in Eq.~(\ref{eq:(M4)}).
In the expression for the dimensionless Jeffreys-Lomnitz creep law, 
it is now convenient to separately consider four cases:
\begin{equation}
\label{eq:(M5bis)}
t\ge 0 \,, \quad \psi(t) = \left\{
\begin{array}{lll}
t/\tau_0\,,  &  \; \alpha = 1\,, \\ \\
{\ds \frac{\left(1 + t/\tau_0\right)^\alpha-1}{\alpha}}\,,&  \; 0<\alpha <1\,, \\ \\
\log(1 +t/\tau_0)\,, &  \; \alpha =0\,, \\ \\
{\ds \frac{1 -\left(1 + t/\tau_0 \right)^{-|\alpha|}}{|\alpha|}}\, &  \; \alpha <0\,.
\end{array}
\right.
\end{equation}
The behaviour of $\psi(t)$ as a function of the dimensionless  time $t/\tau_0$ is
illustrated in Figure~\ref{fig1}, for some values of $\alpha$ in the range $-2 \le \alpha \le 1$, adopting
a logarithmic and a linear time axis in the top and in the bottom frames, respectively.
From Eq.~(\ref{eq:(M5bis)}), the different asymptotic behaviour of $\psi(t)$ as $t \to \infty$
in the cases $0\le \alpha\le 1$ and $\alpha <0$ is apparent. In the former case, $\psi(t)$
diverges  whereas in the latter case it increases up to the finite value $1/|\alpha|$.
\newpage
\newpage
\vsp
In consideration of the asymptotic  behaviour of $\psi(t)$, from the classification of the types 
of viscoelasticity of Table~\ref{table}, we recognize
that the law provided by Eq.~(\ref{eq:(M5bis)}), along with  (\ref{eq:(M1)}),
describes the behavior of a viscoelastic body that shows instantaneous elasticity.
For $\alpha<0$ and $0\le \alpha\le 1$, it falls in  types I and II, respectively.
\vsp
Since the extended Jeffreys-Lomnitz law accounts  for instantaneous elasticity, 
it is possible to write the {\it relaxation modulus} as follows
\begin{equation}
\label{eq:(M2)} 
G(t)= G_0 \phi(t)\,, \quad t\ge 0\,,
\end{equation}
 where $0< G_0=1/J_0<\infty $  and
$\phi(t)$ is a positive decreasing function of time that we refer to as the dimensionless {\it relaxation
function}.
 \vsp
Using Eqs. (\ref{eq:(M1)}) and (\ref{eq:(M2)}) in (\ref{eq:(2.8)}),
the following interrelation in the Laplace domain is found between the dimensionless 
creep and relaxation functions:
   \begin{equation}
\label{eq:(phi-psi)} 
 s\, \widetilde\phi(s)= \frac{1}{1 +qs\,\widetilde \psi(s)}\,,\quad \hbox{or} \quad
 \widetilde\phi(s) = \frac{1/s}{1 +qs\widetilde \psi(s)}\,,
 \end{equation}
 where $\widetilde\phi(s)$ and $\widetilde\psi(s)$ are the Laplace transforms of
 $\phi(t)$ and $\psi(t)$, respectively. 
 From the limiting theorems of Laplace transforms applied to the first of Eqs. (\ref{eq:(phi-psi)}),
 we easily obtain the following:
   \begin{equation}
\label{eq:(phi-limits)} 
\phi(0^+)= 1 \, , \q  \phi(+\infty) =
\left\{
\begin{array}{ll}
 0\,,  & \; 0\le\alpha\le 1\,,\\
 {\ds \frac{1}{1 + q/|\alpha|}}\,, & \; \alpha <0\,.
 \end{array}
 \right.
\end{equation}
However, the analytical derivation of  $\phi(t)$ from inverting the Laplace transforms
in Eqs.   (\ref{eq:(phi-psi)})
  appears as a difficult (if not prohibitive) task even if
  $\widetilde \psi(s)$ 
can be  expressed in terms of  a transcendental function  
related to  incomplete Gamma function, see
\cite{Strick_JGR84}.
  In order to obtain the relaxation function, we numerically solve the Volterra integral
equation of the second kind (\ref{eq:(2.12a)}) with
$J(t)$ appropriate for the extended Jeffreys-Lomnitz law.
After  straightforward algebra,
the Volterra equation reads:
\begin{equation}
\phi(t) = 1 - q \int_0^t \left( 1 + \frac{t-t'}{\tau_{0}}\right)^{\alpha-1}\phi(t')\,dt'\,.
\end{equation}
From now on, for the sake of simplicity, we will assume $q=1$.
The numerical method is implemented in the Fortran routine
\texttt{voltra.for}, available from Numerical Recipes \cite{FORTRAN-RECIPES}.
We note that an alternative semi-analytical method oriented to viscoelasticity  has been first introduced
by Hopkins and Hamming in the late 1950s \cite{Hopkins-Hamming_JAP57,Hopkins-Hamming_JAP58},
and recently revisited by Lin in his book \cite{Lin_BOOK2011}.
We also note that, in the limit $\alpha \to -\infty$, $\phi(t)=1$ and the elastic Hooke model is recovered. 
\vsp
In Figure~\ref{fig2}, we show the   relaxation function $\phi(t)$ versus the dimensionless  time $t/\tau_0$,
 for some values of $\alpha$ in the interval $-2 \le \alpha \le 1$, 
 adopting a logarithmic and a linear time axis
in the top and in the bottom frames, respectively.
\vsp
We conclude this section {with a discussion of} the asymptotic representations
of $\psi(t)$  and $\phi(t)$ as $t/\tau_0 \to 0^+$ and $t/\tau_0 \to +\infty\,,$
which provide simple analytical estimates of the corresponding
functions for sufficiently small and large times, respectively.
These asymptotic results, indeed trivial for $\psi(t)$, are more relevant
for $\phi(t)$, for which we only dispose of  a numerical evaluation.
From the asymptotic analysis we exclude the trivial case $\alpha=1$, for which we
easily recover the creep and relaxation laws of the Maxwell body,
valid for any time:
\begin{equation}
\label{eq:(psi-phi-Maxwell)}
\psi(t)= t/\tau_0\,, \quad \phi(t) =\exp (-t/\tau_0) \quad t\ge 0.
\end{equation}
For the function $\psi(t)$,  the case $t/\tau_0 \to 0^+$ can be easily studied recalling
the binomial  series representation
\begin{equation}
\label{eq:(psi-series)}
\psi(t) = \frac{1}{\alpha}\, \sum_{n=1}^\infty \binom{\alpha}{n}\,
\left(\frac{t}{\tau_0}\right)^n\,,
\end{equation}
which is convergent for $0\le t/\tau_0 <1$. Hence, we obtain the straightforward
asymptotic result
\begin{equation}
\label{eq:(psi-t-zero)}
t/\tau_0 \to 0^+ \,\Longrightarrow \psi(t)\, \sim t/\tau_0 \,,\q  \alpha <1\,.
\end{equation}
In the case $t/\tau_0 \to +\infty$, starting from the definition given by Eq.~(\ref{eq:(M5bis)}),
we have:
\begin{equation}
\label{eq:(psi-t-infty)}
 t/\tau_0 \to +\infty \,\Longrightarrow\,
\psi(t) \sim
\left\{
\begin{array}{lll}
 (t/\tau_0)^\alpha/\alpha\,,  & \; 0<\alpha <1\,, \\  \\
\log(t/\tau_0)\,,&  \;\alpha=0 \,,\\ \\
\left(1 -(t/\tau_0)^{-|\alpha|}\right)/|\alpha|\,,
&  \;\alpha <0\,.
\end{array}
\right.	
\end{equation}
\newpage
\vsp
For the function $\phi(t)$, the asymptotic results for $t/\tau_0 \to 0^+$
and $t/\tau_0 \to +\infty$ can be  obtained by applying the Tauberian theorems 
in the R.H.S. part of Eq. (\ref{eq:(phi-psi)}) (with $q=1$)
 of Laplace transform. 
Because we prefer to avoid the exact expression of $\widetilde \psi(s)$ 
in terms of  transcendental functions,  
 we    will use  the Tauberian theorems in a simplified form. 
This means to invert for
$s\to \infty$ the Laplace transform  
of  $\psi(t)$ as $t \to 0$,  and conversely 
for $s\to 0$ the Laplace transform  
of  $\psi(t)$ as $t \to \infty$.
\vsp
As far the behaviour  as $t/\tau_0 \to 0^+$ is concerned, 
  from Eqs. (\ref{eq:(psi-t-zero)}) and (\ref{eq:(phi-psi)}) we have: 
  $$
  s \to \infty  \,\Longrightarrow \,
  \widetilde{\psi}(s)\sim 1/s^2
  \,\Longrightarrow \,
   \widetilde{\phi}(s)\sim \frac{1/s}{1+ 1/s}
  \sim 1/s -1/s^2\,,
  $$
   so that
    \begin{equation}
	\label{eq:(phi-t-zero)}
	t/\tau_0 \to 0^+ \,\Longrightarrow \,\phi(t)\, \sim 1- t/\tau_0\,, \q  \alpha <1\,.
\end{equation}
As far the behaviour  as $t/\tau_0 \to +\infty$ is concerned,
 from Eqs. (\ref{eq:(psi-t-infty)}) and (\ref{eq:(phi-psi)}) we obtain:
$$
s \to 0 \,\Longrightarrow \,\widetilde{\psi}(s) \sim
\left\{
\begin{array}{lll}
 \Gamma(\alpha)/s^{\alpha+1}\,,  &  \; 0<\alpha <1\,, \\ \\
- (C + \log s)/s\,, &  \;\alpha=0 \,,\\ \\
1/(|\alpha|s)  + \Gamma(-|\alpha|)/ s^{-|\alpha|+1}\,,
&  \;\alpha <0\,,
\end{array}
\right.	
$$
and 
$$ s \to 0 \,  \Longrightarrow \,  \widetilde{\phi}(s)\sim
\left\{
\begin{array}{lll}
{\ds \frac{1/s}{1 +\Gamma(\alpha)/s^{\alpha}}} \sim {\ds\frac{1}{\Gamma(\alpha) s^{1-\alpha}}}\,,
 &  \; 0<\alpha <1\,, \\ \\
 {\ds \frac{1/s}{1 - C - \log s}} \sim  {\ds \frac{1}{s\log (1/s)}}\,,
 & \;\alpha=0 \,,\\ \\
{\ds \frac{1/s}{1+ 1/|\alpha|  + \Gamma(-|\alpha|)/ s^{-|\alpha|}}}\,,
&  \;\alpha <0\,,
\end{array}
\right.	
$$
where for the case $\alpha=0$ the constant $C= -\Gamma^\prime(1)= 0.577215...$ 
denotes the so called Euler-Mascheroni constant.
\newpage
\vsp
 Henceforth
    \begin{equation}
	\label{eq:(phi-t-infty)}
\!\!	t/\tau_0 \to +\infty \Longrightarrow \phi(t) \sim
	\left\{
\begin{array}{lll}
 {\ds   \frac{\sin(\pi\alpha)}{\alpha}} \,  (t/\tau_0)^{-\alpha},   &   0<\alpha <1, \\ \\
{\ds \frac{1}{\log (t/\tau_0)}}, &  \alpha=0 ,\\ \\
{\ds \frac{1}{1+ 1/|\alpha|} \left( 1\!+\! \frac{1}{1+|\alpha|}\, (t/\tau_0)^{-|\alpha|}\right )},
& \alpha <0,
\end{array}
\right.
\end{equation}
\vsp
where the Laplace inversion of $1/(s\log (1/s))$ was obtained via the Karamata-Tauberian
theory of slow varying functions in the paper by Mainardi et al. \cite{Mainardi-et-al_JVC07}.
For details on slow varying functions and  on Karamata-Tauberian theorems, see \textit{e.g.,}
 Feller (1971) \cite{Feller_BOOK71}.

\section{Retardation spectrum for the extended Jeffreys-Lomnitz law of creep}

\begin{figure}[h!]
\begin{center}
\includegraphics[width=0.60\textwidth,angle=-90.]{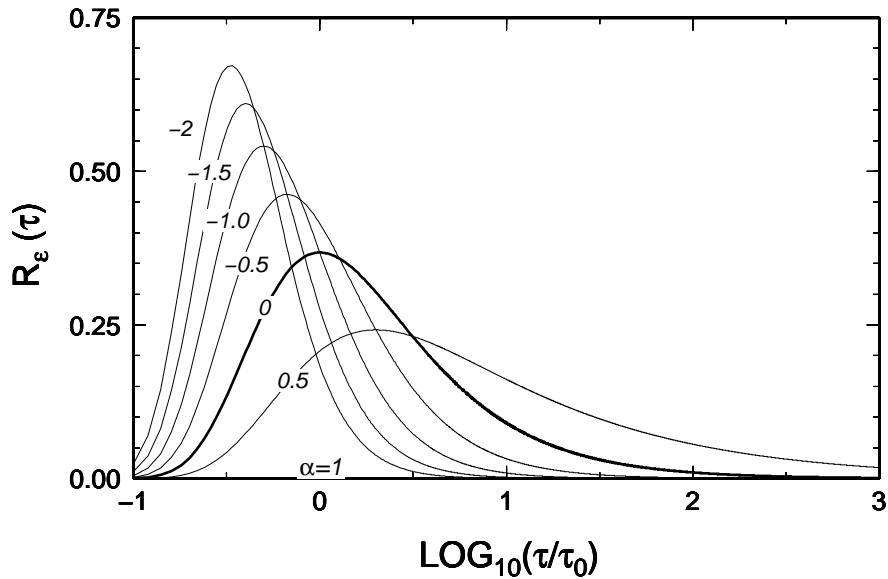}
\caption{
The  retardation spectrum  $R_\epsilon(\tau)$ 
versus dimensionless time $\tau/\tau_0$, for $\alpha = 1, 0.5, 0, -0.5, -1, -1.5, -2$,
adopting a logarithmic scale.}
\label{fig3}
\end{center}
\end{figure}

Following the method outlined in Section 2, see the top Eqs. in  (\ref{eq:(2.34)}) or (\ref{eq:(2.35)}),
we must  consider the derivative
of the dimensionless creep function in the extended Jefferys-Lomnitz law 
as the Laplace transform (with Laplace parameter $\xi =t/\tau_0$) of an
"original function" in the variable $\gamma= \tau_0/t$.
For sake of simplicity, from now on,   we will  scale times and frequencies  by assuming   $\tau_0=1$, 
and also  put the material constants $J_0$ and $q$ equal to 1.
\vsp  
 In order to obtain the required time retardation spectrum we first derive the corresponding  frequency  
 spectrum for $\alpha<1$ by proving the Laplace transform pair:
 \begin{equation}
\label{eq:(M9)} 
\gamma \, S_\epsilon(\gamma) = \frac{1}{\Gamma(1-\alpha)}\, \frac{\e^{-\gamma}}{\gamma^\alpha} 
  = \, \div \,(\xi+1)^{\alpha-1} =
 \frac{d\psi}{d\xi}= \dot J_\tau(\xi)\,,
\q \alpha <1\,.
\end{equation}
The prove is based on the well-known integral representation of the Gamma function,
$$\Gamma(z):= \int_0^\infty \e^{-u} \, u^{(z-1)}\, du\,, \q \textrm{Re}\{z\} >0\,.  $$
In fact 
 $$\frac{1}{\Gamma(1-\alpha)}\,\int_0^\infty \! \e^{-\xi \gamma}\, \frac{\e^{-\gamma}}{\gamma^\alpha}
 =  \frac{(\xi+1)^{\alpha-1}}{\Gamma(1-\alpha)}\, \int_0^\infty \!\e^{-u}\, u^{-\alpha}\, du 
 =(\xi+1)^{\alpha-1}\,$$
where we have used the change of variable $u=(\xi+1)\gamma$ and the integral representation of 
$\Gamma(1-\alpha)$.
Then, from the first expression in Eq. (\ref{eq:(2.32)}) we have
$$  R_\epsilon(1/\gamma)= \gamma^2\, S_\epsilon (\gamma) =
\frac{1}{\Gamma(1-\alpha)}\, \frac{\e^{-\gamma}}{\gamma^{\alpha-1}}\,,
\q \alpha <1\,,
$$
so that, noting $\tau= 1/\gamma$, we derive the required  analytical form for the
retardation spectrum,
\begin{equation}
\label{eq:(M6)} 
R_\epsilon(\tau) =  \frac{1}{\Gamma(1-\alpha)} \, \frac{\e^{-1/\tau}}{\tau^{1-\alpha}}
\,, \q \alpha <1 \,.
\end{equation}
For the limiting values of $\alpha=1$ (Maxwell body) and
$\alpha \to -\infty$ (Hooke body) we recover $R_\epsilon(\tau)=0$  due to the
vanishing of the factor $1/\Gamma(1-\alpha)$.
\vsp
Accounting the previous notes on the asymptotic behavior of $\psi(t)$ 
 we now explore the integrability of such spectrum
on the interval $0<\tau<\infty$. Using directly Eq.~(\ref{eq:(M6)}), we have
\begin{equation}
\label{eq:(M8)} 
\int_0^\infty\!\! R_\epsilon(\tau)\, d\tau = \frac{1}{\Gamma(1-\alpha)}
\int_0^\infty\!\! \frac{\e^{-u}}{u^{\alpha+1}}\, du =
\left\{\begin{array}{ll}
\infty\,,  &\hbox{if}\;   0\le \alpha<1\,,\\
1/|\alpha| &\hbox{if}\;  -\infty<\alpha<0\,.
\end{array}
\right .
\end{equation}
This is obtained by changing the variable $\tau$ into $u=1/\tau$, recalling the integral 
representation of the Gamma function and 
 noting $\Gamma(-\alpha)= -\Gamma(1-\alpha)/\alpha $.
\vsp
Of course the above result 
is consistent with the asymptotic behaviour of our creep law
(\ref{eq:(M5bis)}) as $t/\tau_0 \to +\infty $.
\vsp
In Figure~\ref{fig3}, we show the retardation spectrum  $R_\epsilon(\tau)$
versus the  dimensionless retardation time $\tau/\tau _0$
for some values of $\alpha \in [-2,+1]$, adopting in abscissa  a
logarithmic scale. 
\vsp 
We note that, {for $\alpha=0$ (Lomnitz body), the
maximum of the spectrum occurs just for $\tau/\tau_0=1$. Furthermore,
for $\alpha<0$,} as $\tau \to 0$ the maximum increases with $|\alpha|$
up to infinity in such a way that $R_\epsilon$ identically vanishes in the
open interval $\tau>0$.
\newpage
  \section{Concluding remarks}
By revisiting the Jeffreys-Lomnitz law of creep, we have described
the behaviour of a general viscoelastic body that, according to the value
of a single parameter $\alpha$ ranging from 1 to $-\infty$, shows a transition
from a Maxwell to a Hooke body. 
This continuous transition is well shown in
Figure~\ref{fig1}, where the creep law $\psi(t)$ is displayed for some values
of $\alpha$  in the range $-2\le \alpha\le 1$ for more rheological interest.
We have numerically computed the relaxation modulus $\phi(t)$ corresponding
to our creep law by solving a Volterra integral equation of the second  kind,
see Figure~\ref{fig2}. 
Furthermore, we have provided the analytical expression
of the spectrum of retardation times,
which is shown in Figure~\ref{fig3}.
\vsp
 We are not aware of any previous attempt in the
literature to compute the retardation spectrum and the relaxation law  for the 
Jeffreys-Lomnitz law of creep, so we think to have  filled a gap in this field.
However, we have not numerically evaluated the relaxation spectrum
leaving this problem to interested readers who can chose the most suitable   method to
approximate  the relaxation spectrum from the numerical data of the relaxation modulus.
On  this respect there exists  a huge literature, see e.g.
\cite{Bazant_0,Bazant_1,Bazant_2,Bazant_3,Emri_BOOK2005,Emri-Tschoegl_1,Emri-Tschoegl_4,Emri-Tschoegl_5,
Emri-et-al_JNNFM05,Ferry_BOOK80,Lin_BOOK2011,Park-Schapery_IJSS98,Schapery-Park_IJSS99,
Tschoegl_71b,Tschoegl_BOOK89,Tschoegl_97,Tschoegl-Emri_3,Zi-Bazant_2001}.          
\vsp
We also note that in the recent book by Schilling et al. \cite{Schilling_BOOK10},
where a rigorous mathematical description of Bernstein functions is found,
the authors have considered a function similar to our extended law of creep in two examples
(see No.~2,3 at pp.~218--219, Chapter 15) without the factor $1/\alpha$, corresponding
to our parameter  $0<\alpha <1$ and $-1<\alpha<0$. For these cases they have provided the
so-called {\it L\'evy density} that, as a matter of fact, coincides with our frequency
spectral function $S_\epsilon(\gamma)$ and therefore with our retardation spectrum
$R_\epsilon(\tau)$  by putting $\gamma=1/\tau$. Therefore, in this work, we
have extended their results to the cases $\alpha=0$ and $\alpha \le -1$ by adopting
our notation.
  \vsp
Finally, we   hope that the results obtained for the creep and relaxation laws investigated in this paper may
be  useful for fitting experimental data of creep and/or relaxation responses in rheology of real materials.


\section*{Acknowledgments}
  FM likes to remind his personal
contacts with Sir Harold Jeffreys (in 1973--1974) and with Prof. Ellis Strick (in 1980--1984)
who, in some  way, even if after so many years, have inspired this research work.
GS acknowledges  COST Action ES0701 ``Improved Constraints on Models of Glacial
Isostatic Adjustment''.
We  are grateful to two anonymous reviewers for helpful suggestions.  
The figures have been drawn using the Generic Mapping Tools (GMT) public domain
software \cite{Wessel-Smith_1998}.  

\section*{Appendix: The Post-Widder formula for the retardation spectrum}
The Post-Widder formula provides the original function $f(t)$ from its Laplace transform 
$\widetilde f(s)$, known with all  its  derivatives on the real semi-axis of the complex Laplace plane,
via a limit of infinite sequence as
$$
f(t) = 
{\ds \lim_{n \to \infty} \frac{(-1)^n}{n!} 
\left[ s^{n+1}\, {\widetilde f}^{(n)}(s)\right]_{s={n}/{t}}}
 \eqno(A.1) $$
 where  ${\widetilde f}^{(n)}$ is the $n$--th derivative of ${\widetilde f}$ with respect to the 
Laplace variable $s$.
This is the case when the Laplace transform is proved to be an analytic function  on 
the right-half $s$-plane. For other details we refer e.g. to \cite{Tschoegl_BOOK89}. 
\vsp
The purpose of this Appendix is to prove the validity of   Eq. (\ref{eq:(M9)})
 by using (in a suitable way) the Post-Widder formula. 
In our case, denoting in Eq. (A.1) $t$ by $\gamma$ and $s$ by $\xi$, we must verify
 that
  $$ f(\gamma)= {\ds \lim_{n \to \infty} \frac{(-1)^n}{n!} 
\left[ \xi^{n+1}\, {\widetilde f}^{(n)}(\xi)\right]_{\xi=n/\gamma}}=
   \frac{1}{\Gamma(1-\alpha)} \frac{\e^{-\gamma}}{\gamma^\alpha}\,,
   \eqno(A.2) $$ 
   with ${\widetilde f}(\xi) := (\xi+1)^{\alpha-1}$ and $\alpha< 1$.
Now 
$$ {\widetilde f}^{(n)}(\xi) = (-1)^n\, \frac{\Gamma(n-\alpha+1)}{\Gamma(1-\alpha)}\,
 (\xi + 1)^{\alpha -n -1}\,,  
\q n =1,2, \dots
\eqno(A.3)$$
so we get 
$$
\lim_{n \to \infty} \frac{1}{\Gamma(1-\alpha)}\,
 \frac{\Gamma(n-\alpha +1)}{\Gamma(n+1)}\, \left(\frac{n}{\gamma}\right)^{n+1}\,
\left(1+ \frac{n}{\gamma}\right)^{\alpha-n-1}\,.
 \eqno(A.4) $$
We easily  recognize from Stirling asymptotic formula that for $n \to \infty$ 
$$  \frac{\Gamma(n-\alpha +1)}{\Gamma(n+1)}\sim n^{-\alpha}\,.
 \eqno(A.5) $$
Because
 $$\left(\frac{n}{\gamma}\right)^{n+1}\,\left(1+ \frac{n}{\gamma}\right)^{\alpha-n-1}
= \left(\frac{n}{\gamma}\right)^\alpha \,\left(1 + \frac{\gamma}{n}\right)^{\alpha-n-1}
 \eqno(A.6) $$
we finally get 
$$ f(\gamma) = \frac{1}{\Gamma(1-\alpha)\, \gamma^\alpha}\,
\lim_{n \to \infty} \left(1+ \frac{\gamma}{n}\right)^{-n}\, 
\left(1+ \frac{\gamma}{n}\right)^{\alpha-1} =  
\frac{1}{\Gamma(1-\alpha)} \frac{\e^{-\gamma}}{\gamma^\alpha}\,,
 \eqno(A.7) $$
in view of the Neper limit
$$\lim_{n \to \infty} \left(1+ \frac{1}{n}\right)^{n} = 
\e\,.
 \eqno(A.8) $$
 
\end{document}